\documentstyle[12pt,graphicx,amsmath]{article}
\newlength{\pubnumber} 

\catcode`\@=11
\@addtoreset{equation}{section}

\def\section{\@startsection{section}{1}{\z@}{3.5ex plus 1ex minus .2ex}
 {2.3ex plus .2ex}{\large\bf}}
\def\subsection{\@startsection{subsection}{2}{\z@}{2.3ex plus .2ex}
 {2.3ex plus .2ex}{\bf}}

\newcommand{\ba}{\begin{eqnarray}}
\newcommand{\ea}{\end{eqnarray}}
\begin{document}

\begin{titlepage}
\samepage{
\setcounter{page}{1}
\rightline{LTH--895}
\rightline{April 2011}

\vfill
\begin{center}
 {\Large \bf Top Quark Mass in \\
Exophobic Pati--Salam Heterotic String Model
}
\vspace{1cm}
\vfill {\large
Kyriakos Christodoulides$^{1}$,
Alon E. Faraggi$^{1}$,
 and
John Rizos$^{2}$}\\
\vspace{1cm}
{\it $^{1}$ Dept.\ of Mathematical Sciences,
             University of Liverpool,
         Liverpool L69 7ZL, UK\\}
\vspace{.05in}
{\it $^{2}$ Department of Physics,
              University of Ioannina, GR45110 Ioannina, Greece\\}
\vspace{.025in}
\end{center}
\vfill
\begin{abstract}

We analyse the phenomenology of an exemplary exophobic
Pati--Salam heterotic string vacuum, in which
no exotic fractionally charged states exist in the massless string spectrum.
Our model also contains the
Higgs representations that are needed to break the gauge symmetry
to that of the Standard Model and to generate fermion masses
at the electroweak scale.
We show that the requirement of a leading mass term for the
heavy generation,  which is not degenerate with the mass
terms of the lighter generations, places
an additional strong constraint on the viability of the models.
In many models a top quark Yukawa may not exist at all, whereas
in others two or more generations may obtain a mass term at leading
order. In our exemplary model  a mass term at leading
order exist only for one family. Additionally,
we demonstrate the existence of supersymmetric $F$-- and $D$--flat
directions that give heavy mass to all the
colour triplets beyond those of the Standard Model
and leave one pair of electroweak Higgs doublets light.
Hence, below the Pati--Salam breaking scale, the matter states in our
model that are charged under the observable gauge symmetries,
consist solely of those of the Minimal Supersymmetric Standard Model.

\noindent

\end{abstract}
\smallskip}
\end{titlepage}

\setcounter{footnote}{0}

\def\beq{\begin{equation}}
\def\eeq{\end{equation}}
\def\beqn{\begin{eqnarray}}
\def\eeqn{\end{eqnarray}}

\def\no{\noindent }
\def\nolabel{\nonumber }
\def\ie{{\it i.e.}}
\def\eg{{\it e.g.}}
\def\half{{\textstyle{1\over 2}}}
\def\third{{\textstyle {1\over3}}}
\def\quarter{{\textstyle {1\over4}}}
\def\sixth{{\textstyle {1\over6}}}
\def\m{{\tt -}}
\def\p{{\tt +}}

\def\Tr{{\rm Tr}\, }
\def\tr{{\rm tr}\, }

\def\slash#1{#1\hskip-6pt/\hskip6pt}
\def\slk{\slash{k}}
\def\GeV{\,{\rm GeV}}
\def\TeV{\,{\rm TeV}}
\def\y{\,{\rm y}}
\def\SM{Standard--Model }
\def\SUSY{supersymmetry }
\def\SSSM{supersymmetric standard model}
\def\vev#1{\left\langle #1\right\rangle}
\def\l{\langle}
\def\r{\rangle}
\def\o#1{\frac{1}{#1}}

\def\Htw{{\tilde H}}
\def\chibar{{\overline{\chi}}}
\def\qbar{{\overline{q}}}
\def\ibar{{\overline{\imath}}}
\def\jbar{{\overline{\jmath}}}
\def\Hbar{{\overline{H}}}
\def\Qbar{{\overline{Q}}}
\def\abar{{\overline{a}}}
\def\alphabar{{\overline{\alpha}}}
\def\betabar{{\overline{\beta}}}
\def\tautwo{{ \tau_2 }}
\def\thetatwo{{ \vartheta_2 }}
\def\thetathree{{ \vartheta_3 }}
\def\thetafour{{ \vartheta_4 }}
\def\ttwo{{\vartheta_2}}
\def\tthree{{\vartheta_3}}
\def\tfour{{\vartheta_4}}
\def\ti{{\vartheta_i}}
\def\tj{{\vartheta_j}}
\def\tk{{\vartheta_k}}
\def\calF{{\cal F}}
\def\smallmatrix#1#2#3#4{{ {{#1}~{#2}\choose{#3}~{#4}} }}
\def\ab{{\alpha\beta}}
\def\Minv{{ (M^{-1}_\ab)_{ij} }}
\def\bone{{\bf 1}}
\def\ii{{(i)}}
\def\V{{\bf V}}
\def\N{{\bf N}}

\def\b{{\bf b}}
\def\S{{\bf S}}
\def\X{{\bf X}}
\def\I{{\bf I}}
\def\mb{{\mathbf b}}
\def\mS{{\mathbf S}}
\def\mX{{\mathbf X}}
\def\mI{{\mathbf I}}
\def\balpha{{\mathbf \alpha}}
\def\bbeta{{\mathbf \beta}}
\def\bgamma{{\mathbf \gamma}}
\def\bxi{{\mathbf \xi}}

\def\t#1#2{{ \Theta\left\lbrack \matrix{ {#1}\cr {#2}\cr }\right\rbrack }}
\def\C#1#2{{ C\left\lbrack \matrix{ {#1}\cr {#2}\cr }\right\rbrack }}
\def\tp#1#2{{ \Theta'\left\lbrack \matrix{ {#1}\cr {#2}\cr }\right\rbrack }}
\def\tpp#1#2{{ \Theta''\left\lbrack \matrix{ {#1}\cr {#2}\cr }\right\rbrack }}
\def\l{\langle}
\def\r{\rangle}
\newcommand{\cc}[2]{c{#1\atopwithdelims[]#2}}
\newcommand{\nn}{\nonumber}


\def\inbar{\,\vrule height1.5ex width.4pt depth0pt}

\def\IC{\relax\hbox{$\inbar\kern-.3em{\rm C}$}}
\def\IQ{\relax\hbox{$\inbar\kern-.3em{\rm Q}$}}
\def\IR{\relax{\rm I\kern-.18em R}}
 \font\cmss=cmss10 \font\cmsss=cmss10 at 7pt
\def\IZ{\relax\ifmmode\mathchoice
 {\hbox{\cmss Z\kern-.4em Z}}{\hbox{\cmss Z\kern-.4em Z}}
 {\lower.9pt\hbox{\cmsss Z\kern-.4em Z}}
 {\lower1.2pt\hbox{\cmsss Z\kern-.4em Z}}\else{\cmss Z\kern-.4em Z}\fi}

\def\AEF{A.E. Faraggi}
\def\JHEP#1#2#3{{\it JHEP}\/ {\bf #1} (#2) #3}
\def\NPB#1#2#3{{\it Nucl.\ Phys.}\/ {\bf B#1} (#2) #3}
\def\PLB#1#2#3{{\it Phys.\ Lett.}\/ {\bf B#1} (#2) #3}
\def\PRD#1#2#3{{\it Phys.\ Rev.}\/ {\bf D#1} (#2) #3}
\def\PRL#1#2#3{{\it Phys.\ Rev.\ Lett.}\/ {\bf #1} (#2) #3}
\def\PRT#1#2#3{{\it Phys.\ Rep.}\/ {\bf#1} (#2) #3}
\def\MODA#1#2#3{{\it Mod.\ Phys.\ Lett.}\/ {\bf A#1} (#2) #3}
\def\IJMP#1#2#3{{\it Int.\ J.\ Mod.\ Phys.}\/ {\bf A#1} (#2) #3}
\def\nuvc#1#2#3{{\it Nuovo Cimento}\/ {\bf #1A} (#2) #3}
\def\RPP#1#2#3{{\it Rept.\ Prog.\ Phys.}\/ {\bf #1} (#2) #3}
\def\EJP#1#2#3{{\it Eur.\ Phys.\ Jour.}\/ {\bf C#1} (#2) #3}
\def\etal{{\it et al\/}}

\hyphenation{su-per-sym-met-ric non-su-per-sym-met-ric}
\hyphenation{space-time-super-sym-met-ric}
\hyphenation{mod-u-lar mod-u-lar--in-var-i-ant}


\setcounter{footnote}{0}
\section{Introduction}

The Standard Model of particle physics remains unscathed by contemporary
experiments. Its augmentation with the right--handed neutrinos, as envisioned
by Pati and Salam nearly four decades ago
\cite{ps}, is mandated by solar and terrestrial
neutrino observations. The Pati--Salam model naturally leads to the
embedding of the standard model in $SO(10)$ representations. Most
strikingly the matter embedding in three 16 spinorial representations
correlates the 54 gauge charges of the Standard Model states into
the single number of spinorial multiplets. The reduction
in the number of experimental parameters from fifty four to one
provides the most important clue for the fundamental origins of the
Standard Model. The remaining parameters, and in particular the
flavour parameters, must find their origin in a theory that
unifies gauge theories with gravity. It is then of further appeal
that heterotic--string theory accommodates
the $SO(10)$ embedding of the Standard Model matter spectrum.
Three generation Heterotic--string models that preserve the
$SO(10)$ embedding of the Standard Model states were constructed
since the late eighties \cite{revamp,fny,alr,eu,cfn,lrs}.

Absence of higher order Higgs representations in heterotic--string models
that are based on level one Kac--Moody current algebras necessitates
that the $SO(10)$
symmetry is broken directly at the string level by discrete Wilson lines.
A well known theorem due to Schellekens \cite{schellekens} states that
any such string model that preserves the
canonical $SO(10)$--GUT embedding
of the weak hypercharge, and in which the non--Abelian GUT symmetries are
broken by discrete Wilson lines, necessarily contain states that carry
charges that do not obey the original GUT quantisation rule
\cite{schellekens}\footnote{
A similar observation was made in the context of Calabi--Yau compactification
models with $E_6$ gauge group broken by Wilson lines \cite{ww}.}.
In terms of the Standard Model charges these exotic states carry
fractional electric charge. Electric charge conservation implies that the
lightest of these states is stable, and their existence in nature
is severely constrained by experiments \cite{halyo}.

While the existence of fractionally charged states in string models that
preserve the canonical $SO(10)$ embedding of the Standard Model
states, and in which the $SO(10)$ symmetry is broken by Wilson lines,
is mandated by Schellekens theorem, they may appear only in vector--like
representations, rather than in chiral representations. Superpotential
terms for the vector--like states can then generate an intermediate
or string scale
mass to the exotic states, through the VEVs of Standard Model singlet fields
\cite{fc,cfn}.
However, as the generation of the VEVs is obtained in an effective field theory
analysis a more appealing solution is to find string models in which the
exotic fractionally charged states are confined to the massive
spectrum. Recently, we demonstrated the existence of Pati--Salam vacua
in which exotic fractionally charged
states do not exist in the massless spectrum \cite{acfkr}. We dubbed
such models as exophobic string vacua. We further showed that
there exist such exophobic Pati--Salam string models that contain
three generations and the required Higgs states to produce
realistic mass spectrum.
We demonstrated the existence of exophobic string vacua by utilising the
free fermionic classification techniques. These methods were developed in ref.
\cite{gkr} for type II string $N=2$ supersymmetric vacua.
They were extended in refs.
\cite{fknr,fkr} for the classification of heterotic $Z_2\times Z_2$ free
fermionic orbifolds, with unbroken $SO(10)$ and $E_6$ GUT symmetries,
and in ref. \cite{acfkr} heterotic--string vacua in which the $SO(10)$ symmetry
is broken to the Pati--Salam subgroup.

The classification method used in
refs. \cite{gkr,fknr, fkr, acfkr} utilises symmetric boundary conditions
for the set of internal world--sheet fermions that correspond to the
six dimensional compactified lattice. The symmetric boundary conditions
correspond to $Z_2$ shifts in the compactified six dimensional torus
and enable the scan of large sets of vacua. Such symmetric assignments in
Pati--Salam heterotic string models lead to the projection of the
untwisted Higgs bi--doublets and preservation of the corresponding
colour triplets \cite{dts}. In quasi--realistic free fermionic models
untwisted Higgs doublets couple to twisted matter states.
The leading coupling is identified with the top quark mass
term in the superpotential \cite{topyuk}.
Hence, this coupling is not present
in the exophobic Pati--Salam models of ref. \cite{acfkr}.
The question arises whether a top quark mass term exists
in these string vacua. A viable
top quark Yukawa term is one of the first criteria
that a realistic string vacuum should admit.

An alternative to the twisted--twisted--untwisted coupling that
is used in the quasi--realistic free fermionic models is
a twisted--twisted--twisted coupling. The existence of
a viable coupling is model dependent. The three states appearing
in the trilevel term must arise from the three distinct twisted
sectors. Hence, for example, if all the vectorial and spinorial
twisted states would arise from a single sector, the vacuum would
not be viable. In this paper we examine this question in the
exophobic string vacuum of ref. \cite{acfkr}. We show
in one concrete model that the required coupling does
exist. Additionally, we calculate the entire cubic level superpotential
and show the existence of flat directions that leave a light pair
of electroweak Higgs doublets and give heavy mass to all vector--like
colour triplets. Hence, below the Pati--Salam breaking scale
the spectrum of our model coincides with that of the Minimal
Supersymmetric Standard Model (MSSM).


\section{Exophobic Pati--Salam Heterotic--String Model}\label{model}

Our exophobic Pati--Salam heterotic--string model is constructed in the
free fermionic formulation \cite{fff}.
In this formulation a string model is
specified in terms of a set of boundary condition basis vectors
$v_i,i=1,\dots,N$
$$v_i=\left\{\alpha_i(f_1),\alpha_i(f_{2}),\alpha_i(f_{3}))\dots\right\},$$
for the 64 world--sheet real fermions \cite{fff},
and the one--loop Generalised GGSO projection coefficients,
$ \cc{v_i}{v_j}.$
The basis vectors span a space $\Xi$ which consists of $2^N$ sectors that give
rise to the string spectrum. Each sector, $\eta\in \Xi$, is given by
\begin{equation}
\eta = \sum N_i v_i,\ \  N_i =0,1
\end{equation}
The spectrum is truncated by a generalised GSO projection whose action on a
string state  $|S>$ is
\begin{equation}\label{eq:gso}
e^{i\pi v_i\cdot F_S} |S> = \delta_{S}\ \cc{S}{v_i} |S>,
\end{equation}
where $F_S$ is the fermion number operator and $\delta_{S}=\pm1$ is the
space--time spin statistics index.
The world--sheet free fermions in the light-cone gauge in the
usual notation are:
$\psi^\mu, \chi^i,y^i, \omega^i, i=1,\dots,6$ (left-movers) and
$\bar{y}^i,\bar{\omega}^i, i=1,\dots,6$,
$\psi^A, A=1,\dots,5$, $\bar{\eta}^B, B=1,2,3$, $\bar{\phi}^\alpha,
\alpha=1,\ldots,8$ (right-movers).
The exophobic Pati--Salam model is
generated by a set of thirteen basis vectors
$
B=\{v_1,v_2,\dots,v_{13}\},
$
where
\begin{eqnarray}
v_1=1&=&\{\psi^\mu,\
\chi^{1,\dots,6},y^{1,\dots,6}, \omega^{1,\dots,6}| \nonumber\\
& & ~~~\bar{y}^{1,\dots,6},\bar{\omega}^{1,\dots,6},
\bar{\eta}^{1,2,3},
\bar{\psi}^{1,\dots,5},\bar{\phi}^{1,\dots,8}\},\nonumber\\
v_2=S&=&\{\psi^\mu,\chi^{1,\dots,6}\},\nonumber\\
v_{2+i}=e_i&=&\{y^{i},\omega^{i}|\bar{y}^i,\bar{\omega}^i\}, \
i=1,\dots,6,\nonumber\\
v_{9}=b_1&=&\{\chi^{34},\chi^{56},y^{34},y^{56}|\bar{y}^{34},
\bar{y}^{56},\bar{\eta}^1,\bar{\psi}^{1,\dots,5}\},\label{basis}\\
v_{10}=b_2&=&\{\chi^{12},\chi^{56},y^{12},y^{56}|\bar{y}^{12},
\bar{y}^{56},\bar{\eta}^2,\bar{\psi}^{1,\dots,5}\},\nonumber\\
v_{11}=z_1&=&\{\bar{\phi}^{1,\dots,4}\},\nonumber\\
v_{12}=z_2&=&\{\bar{\phi}^{5,\dots,8}\},\nonumber\\
v_{13}=\alpha &=& \{\bar{\psi}^{4,5},\bar{\phi}^{1,2}\}.\nonumber
\end{eqnarray}
The first two basis vectors generate a model with $N=4$
space--time supersymmetry and $SO(44)$ gauge group in four
dimensions. The next six basis vectors correspond to freely
acting shifts on the internal six dimensional compactified torus
and reduce the gauge symmetry to $SO(32)$. The basis vectors $z_1$
and $z_2$ are freely acting as well, and reduce the gauge symmetry
arising from the Neveu--Schwarz (NS) sector to
$SO(16)\times SO(8)\times SO(8)$. Additional space--times
vector bosons may arise from the sectors \cite{fknr,fkr,acfkr}
\begin{equation}
\mathbf{G} =
\left\{ \begin{array}{cccccc}
z_1          ,&
z_2          ,&
\alpha       ,&
\alpha + z_1 ,&
              &
                \cr
x            ,&
z_1 + z_2    ,&
\alpha + z_2 ,&
\alpha + z_1 + z_2,&
\alpha + x ,&
\alpha + x + z_1
\end{array} \right\} \label{stvsectors}
\end{equation}
and enhance the four dimensional gauge group. In (\ref{stvsectors})
we defined the vector combination $$x=1+S+\sum_{i=1}^6 e_i+z_1+z_2,$$
which may enhance the observable $SO(16)$ gauge symmetry to $E_8$.
For suitable choices of the GGSO projection coefficients all
the space--time vector bosons arising from the sectors in eq.
(\ref{stvsectors}) are projected out.
The basis vectors $b_1$ and $b_2$ correspond to the
$Z_2\times Z_2$ twists of a $Z_2\times Z_2$ orbifold.
Each $Z_2$ twist reduces the number of supersymmetry generators
from $N=4$ to $N=2$. In combination $b_1$ and $b_2$
break $N=4$ to $N=1$ space--time supersymmetry,
and reduce the NS gauge symmetry to
$SO(10)\times U(1)^3\times SO(8)\times SO(8)$.

In the quasi--realistic heterotic string models the gauge symmetries
are realised as level one Kac--Moody algebras. The
massless spectrum of such models
does not contain scalar Higgs multiplets in the adjoint
representation that can be used to break the non--Abelian $SO(10)$
GUT symmetry. Consequently, the GUT gauge group must be broken
at the string level, by a boundary condition basis vector in the
free fermionic formalism, or a discrete Wilson line in the orbifold
formalism. The basis vector $\alpha$ reduces the $SO(10)$ symmetry to
the Pati--Salam subgroup. The gauge group in our model is therefore:
\beqn
{\rm observable} ~: &~~~~~~~~SO(6)\times SO(4) \times U(1)^3 \nonumber\\
{\rm hidden}     ~: &~~SO(4)^2\times SO(8)~~~~             \nonumber
\eeqn
The matter states in our model are embedded in $SU(4)\times{SU(2)}_L\times{SU(2)}_R$
representations as follows:
\begin{align}
    {F}_L\left({\bf4},{\bf2},{\bf1}\right)     &\rightarrow
     q\left({\bf3},{\bf2},-\frac 16\right) + \ell{\left({\bf1},{\bf2},\frac 12\right)}
     \nonumber\\
\bar{F}_R\left({\bf\bar 4},{\bf1},{\bf2}\right)&\rightarrow   u^c\left({\bf\bar 3},
{\bf1},\frac 23\right)+d^c\left({\bf\bar 3},{\bf1},-\frac 13\right)+
                            e^c\left({\bf1},
                           {\bf1},-1)+\nu^c({\bf1},{\bf1},0\right)
                           \nonumber\\
h({\bf1},{\bf2},{\bf2})
 &\rightarrow   h^d\left({\bf1},{\bf2},\frac 12\right) + h^u\left({\bf1},{\bf2},-\frac 12\right)
 \nonumber\\
D\left({\bf6},{\bf1},1\right)  &\rightarrow   d_3\left({\bf3},{\bf1},\frac 13\right) +
\bar{d}_3\left({\bf\bar 3},{\bf1},-\frac 13\right),
                             \nonumber
\end{align}
where $F_L$ and ${\bar F}_R$ contain a single Standard Model generation;
$h^d$ and $h^u$ are electroweak Higgs doublets; and $D$ contains
vector--like colour triplets. The decomposition of the
Pati--Salam breaking Higgs fields in terms of the Standard Model group factors
is:
\begin{align}
\bar{H}({\bf\bar 4},{\bf1},{\bf2})&\rightarrow   u^c_H\left({\bf\bar 3},
{\bf1},\frac 23\right)+d^c_H\left({\bf\bar 3},{\bf1},-\frac 13\right)+
                            \nu^c_H\left({\bf1},{\bf1},0\right)+
                             e^c_H\left({\bf1},{\bf1},-1\right)
                             \nonumber\\
{H}\left({\bf4},{\bf1},{\bf2}\right)&\rightarrow  u_H\left({\bf3},{\bf1},-\frac
23\right)+d_H\left({\bf3},{\bf1},\frac 13\right)+
              \nu_H\left({\bf1},{\bf1},0\right)+ e_H\left({\bf1},{\bf1},1\right)\nn
\end{align}
The electric charge in the Pati--Salam models is given by:
\beq
Q_{em} = {1\over\sqrt{6}}T_{15}+{1\over2}I_{3_L}+{1\over2}I_{3_R}
\eeq
where $T_{15}$ is the diagonal generator of $SU(4)$ and
$I_{3_L}$, $I_{3_R}$
are the diagonal generators of $SU(2)_L$, $SU(2)_R$, respectively.

The second ingredient that is needed to define the string vacuum
are the GGSO projection coefficients that appear in the
one--loop partition function,
$\cc{v_i}{v_j}$, spanning a $13\times 13$ matrix.
Only the elements with $i>j$ are
independent, and the others are fixed by modular invariance.
A priori there are therefore 78 independent coefficients corresponding
to $2^{78}$ distinct string vacua. Eleven coefficients
are fixed by requiring that the models possess $N=1$ supersymmetry.
Additionally, imposing the
condition that the only space--time vector bosons that
remain in the spectrum are those
that arise from the untwisted sector restricts the number of phases
to a total of 51 independent GGSO phases. Each distinct configuration
of these phases corresponds to a distinct vacuum. Some degeneracy in this
space of models may still exist due to additional symmetries over the entire
space. This is not relevant for our purposes here as our aim in this
work is to extract from the total space an exemplary model with the
required phenomenological properties. Statistical analysis over the
entire space was presented in ref. \cite{acfkr}.

The breaking of the $SO(10)$ GUT symmetry by the $\alpha$ boundary
condition basis vector results in combinations of the basis vectors
that can produce a priori massless states with fractional electric
charge. All these sectors, and the type of states that they a priori
can give rise to, are enumerated in ref. \cite{acfkr}.

By employing an algorithm
to generate random selection of the GGSO projection coefficient
the Pati--Salam free fermionic heterotic--string vacua were
classified in ref. \cite{acfkr}.
For suitable choices of the GGSO projection coefficients
all the massless
fractionally charged states are projected out.
Fractionally charged states
in this case only exist in the massive string spectrum, which is
compatible with experimental constraints.
An explicit choice of GGSO projection coefficients that produces a
model with this property is given by:

\beq \label{BigMatrix}  (v_i|v_j)\ \ =\ \ \bordermatrix{
      & 1& S&e_1&e_2&e_3&e_4&e_5&e_6&b_1&b_2&z_1&z_2&\alpha\cr
 1    & 1& 1& 1& 1& 1& 1& 1& 1& 1& 1& 1& 1& 0\cr
S     & 1& 1& 1& 1& 1& 1& 1& 1& 1& 1& 1& 1& 1\cr
e_1   & 1& 1& 0& 0& 0& 0& 0& 0& 0& 0& 0& 0& 1\cr
e_2   & 1& 1& 0& 0& 1& 0& 0& 1& 0& 0& 1& 1& 0\cr
e_3   & 1& 1& 0& 1& 0& 1& 1& 0& 0& 0& 1& 0& 0\cr
e_4   & 1& 1& 0& 0& 1& 0& 1& 0& 0& 0& 1& 0& 1\cr
e_5   & 1& 1& 0& 0& 1& 1& 0& 0& 1& 0& 1& 1& 1\cr
e_6   & 1& 1& 0& 1& 0& 0& 0& 0& 0& 1& 1& 0& 0\cr
b_1   & 1& 0& 0& 0& 0& 0& 1& 0& 1& 0& 0& 1& 0\cr
b_2   & 1& 0& 0& 0& 0& 0& 0& 1& 0& 1& 1& 0& 0\cr
z_1   & 1& 1& 0& 1& 1& 1& 1& 1& 0& 1& 1& 1& 1\cr
z_2   & 1& 1& 0& 1& 0& 0& 1& 0& 1& 0& 1& 1& 1\cr
\alpha& 0& 1& 1& 0& 0& 1& 1& 0& 1& 1& 0& 1& 0\cr
  }
\eeq
where we introduced the notation
$\cc{v_i}{v_j} = e^{i\pi (v_i|v_j)}$.

\begin{table}[!h]
\noindent
{\small
\openup\jot
\begin{tabular}{|l|l|c|c|c|c|}
\hline
sector&field&$SU(4)\times{SU(2)}_L\times{SU(2)}_R$&${U(1)}_1$&${U(1)}_2$&${U(1)}_3$\\
\hline
$S$&$D_1$&$(6,1,1)$&$+1$&$\hphantom{+}0$&$\hphantom{+}0$\\
&$D_2$&$(6,1,1)$&$\hphantom{+}0$&$+1$&$\hphantom{+}0$\\
&$D_3$&$(6,1,1)$&$\hphantom{+}0$&$\hphantom{+}0$&$+1$\\
&$\bar{D}_1$&$(6,1,1)$&$-1$&$\hphantom{+}0$&$\hphantom{+}0$\\
&$\bar{D}_2$&$(6,1,1)$&$\hphantom{+}0$&$-1$&$\hphantom{+}0$\\
&$\bar{D}_3$&$(6,1,1)$&$\hphantom{+}0$&$\hphantom{+}0$&$-1$\\
&$\Phi_{12}$&$(1,1,1)$&$+1$&$+1$&$\hphantom{+}0$\\
&$\Phi_{12}^{-}$&$(1,1,1)$&$+1$&$-1$&$\hphantom{+}0$\\
&$\bar{\Phi}_{12}$&$(1,1,1)$&$-1$&$-1$&$\hphantom{+}0$\\
&$\bar{\Phi}_{12}^{-}$&$(1,1,1)$&$-1$&$+1$&$\hphantom{+}0$\\
&$\Phi_{13}$&$(1,1,1)$&$+1$&$\hphantom{+}0$&$+1$\\
&$\Phi_{13}^-$&$(1,1,1)$&$+1$&$\hphantom{+}0$&$-1$\\
&$\bar{\Phi}_{13}$&$(1,1,1)$&$-1$&$\hphantom{+}0$&$-1$\\
&$\bar{\Phi}_{13}^-$&$(1,1,1)$&$-1$&$\hphantom{+}0$&$+1$\\
&$\Phi_i,i=1,\dots,6$&$(1,1,1)$&$\hphantom{+}0$&$\hphantom{+}0$&$\hphantom{+}0$\\
&$\Phi_{23}$&$(1,1,1)$&$\hphantom{+}0$&$+1$&$+1$\\
&$\Phi_{23}^-$&$(1,1,1)$&$\hphantom{+}0$&$+1$&$-1$\\
&$\bar{\Phi}_{23}$&$(1,1,1)$&$\hphantom{+}0$&$-1$&$-1$\\
&$\bar{\Phi}_{23}^-$&$(1,1,1)$&$\hphantom{+}0$&$-1$&$+1$\\
\hline
\end{tabular}
}
\caption{\label{tablea}\it
Untwisted matter spectrum and
$SU(4)\times{SU(2)}_L\times{SU(2)}_R\times{U(1)}^3$ quantum numbers. }
\end{table}

The twisted massless states generated in the string vacuum
of eq. (\ref{BigMatrix}) produce the needed spectrum for viable
phenomenology.
It contains three chiral generations; one pair of heavy Higgs
states to break the Pati--Salam gauge symmetry along a flat direction;
light Higgs bi-doublets needed to break the electroweak
symmetry and generate fermion masses; one vector sextet of $SO(6)$ needed
for the missing partner mechanism; it is completely free of massless
exotic fractionally charged states.
States in vectorial representation are obtained in the free fermionic
models from the untwisted Neveu--Schwarz sector and from twisted sectors that
contain four periodic world--sheet right--moving complex fermions.
Massless states
are obtained in such sectors by acting on the vacuum with a Neveu--Schwarz
right--moving fermionic oscillator.
The model of eq. (\ref{BigMatrix})
contains three pairs of untwisted $SO(6)$ sextets, and an additional
sextet from a twisted sector.
These can obtain string scale mass along flat directions.
Additionally, it contains a number of $SO(10)$ singlet states,
some of which transform in non--trivial representations of the
hidden sector gauge group.
The full massless spectrum of the model is shown in tables
\ref{tablea}, \ref{tableb} and \ref{tablec}, where we define
the vector combination $b_3\equiv b_1+b_2+x$.

\begin{table}[!h]
\noindent
{\small
\openup\jot
\begin{tabular}{|l|l|c|c|c|c|}
\hline
sector&field&$SU(4)\times{SU(2)}_L\times{SU(2)}_R$&${U(1)}_1$&${U(1)}_2$&${U(1)}_3$\\
\hline
$S+b_2+e_1+e_6$&${F}_{1L}$&$({4},2,1)$&$\hphantom{+}0$&$-{1/2}$&$\hphantom{+}0$\\
$S+b_2+e_6$&$\bar{F}_{1R}$&$(\bar{4},1,2)$&$\hphantom{+}0$&$-{1/2}$&$\hphantom{+}0$\\
$S+b_3+e_1+e_2+e_3$&${F}_{2L}$&$({4},2,1)$&$\hphantom{+}0$&$\hphantom{+}0$&$-{1/2}$\\
$S+b_1+e_4+e_5$&$\bar{F}_{2R}$&$(\bar{4},1,2)$&$\hphantom{+}{1/2}$&$\hphantom{+}0$&$\hphantom{+}0$\\
$S+b_1+e_3+e_4+e_5+e_6$&${F}_{1R}$&$({4},1,2)$&$-{1/2}$&$\hphantom{+}0$&$\hphantom{+}0$\\
$S+b_3+e_1+e_2+e_4$&$\bar{F}_{3R}$&$(\bar{4},1,2)$&$\hphantom{+}0$&$\hphantom{+}0$&$\hphantom{-}{1/2}$\\
$S+b_3+e_2+e_4$&${F}_{3L}$&$({4},2,1)$&$\hphantom{+}0$&$\hphantom{+}0$&$\hphantom{+}{1/2}$\\
$S+b_3+e_2+e_3$&$\bar{F}_{4R}$&$(\bar{4},1,2)$&$\hphantom{+}0$&$\hphantom{+}0$&$-{1/2}$\\
\hline
$S+b_2+x+e_2+e_5$&$h_1$&$(1,2,2)$&$-{1/2}$&$\hphantom{+}0$&$-{1/2}$\\
$S+b_1+x+e_3+e_5$&$h_2$&$(1,2,2)$&$\hphantom{+}{1/2}$&$\hphantom{+}0$&$\hphantom{+}{1/2}$\\
$S+b_1+x+e_3+e_5+e_6$&$h_3$&$(1,2,2)$&$\hphantom{+}0$&$\hphantom{+}{1/2}$&$\hphantom{+}{1/2}$\\
\hline
$S+b_3+x+e_2$&$\zeta_1$&$(1,1,1)$&$\hphantom{+}{1/2}$&$-{1/2}$&$\hphantom{+}0$\\
$$&$\bar{\zeta}_1$&$(1,1,1)$&$-{1/2}$&$\hphantom{+}{1/2}$&$\hphantom{+}0$\\
\hline
$S+b_3+x+e_1+e_2+e_3+e_4$&$\zeta_2$&$(1,1,1)$&$\hphantom{+}{1/2}$&$\hphantom{+}{1/2}$&$\hphantom{+}0$\\
$ $&$\bar{\zeta}_2$&$(1,1,1)$&$-{1/2}$&$-{1/2}$&$\hphantom{+}0$\\
\hline
$S+b_2+x+e_1+e_2+e_5$&$D_4$&$(6,1,1)$&$-{1/2}$&$\hphantom{+}0$&$-{1/2}$\\
&$\zeta_a, a=3,4$&$(1,1,1)$&$\hphantom{+}{1/2}$&$\hphantom{+}0$&$-{1/2}$\\
&$\bar{\zeta}_a, a=3,4$&$(1,1,1)$&$-{1/2}$&$\hphantom{+}0$&$\hphantom{+}{1/2}$\\
&$\chi_+$&$(1,1,1)$&$\hphantom{+}{1/2}$&$\hphantom{+}{1/2}$&$\hphantom{+}1$\\
&${\chi}_-$&$(1,1,1)$&$\hphantom{+}{1/2}$&$\hphantom{+}{1/2}$&$-1$\\
\hline
$S+b_1+x+e_3+e_4+e_5$&$      \zeta_5$&$(1,1,1)$&$\hphantom{+}0$&$\hphantom{+}{1/2}$&$\hphantom{+}{1/2}$\\
$                  $&$\bar{\zeta}_5$&$(1,1,1)$&$\hphantom{+}0$&$-{1/2}$&$-{1/2}$\\
\hline
$S+b_1+x+e_4+e_5+e_6$&$      \zeta_6$&$(1,1,1)$&$\hphantom{+}0$&$\hphantom{+}{1/2}$&$\hphantom{+}{1/2}$\\
$                  $&$\bar{\zeta}_6$&$(1,1,1)$&$\hphantom{+}0$&$-{1/2}$&$-{1/2}$\\
\hline
$S+b_2+x$&$      \zeta_7$&$(1,1,1)$&$\hphantom{+}{1/2}$&$\hphantom{+}0$&$-{1/2}$\\
$                  $&$\bar{\zeta}_7$&$(1,1,1)$&$-{1/2}$&$\hphantom{+}0$&$\hphantom{+}{1/2}$\\
\hline
\end{tabular}
}
\caption{\label{tableb}\it
Twisted matter spectrum (observable sector)  and
$SU(4)\times{SU(2)}_L\times{SU(2)}_R\times{U(1)}^3$ quantum numbers. }
\end{table}

\begin{table}
\noindent
{\small
\begin{tabular}{|l|l|c|c|c|c|}
\hline
sector&field&${SU(2)}^4\times{SO(8)}$&${U(1)}_1$&${U(1)}_2$&${U(1)}_3$\\
\hline
$S+b_3+x+e_1+e_4$&$H_{12}^1$&$(2,2,1,1,1)$&$-{1/2}$&$-{1/2}$&$\hphantom{+}0$\\\hline
$S+b_3+x+e_1+e_2+e_3$&$H_{12}^2$&$(2,2,1,1,1)$&$\hphantom{+}{1/2}$&$-{1/2}$&$\hphantom{+}0$\\\hline
$S+b_2+x+e_2+e_5+e_6$&$H_{12}^3$&$(2,1,2,1,1)$&$\hphantom{+}{1/2}$&$\hphantom{+}0$&$-{1/2}$\\\hline
$S+b_3+x+e_2+e_3$&$H_{34}^1$&$(1,1,2,2,1)$&$\hphantom{+}{1/2}$&$-{1/2}$&$\hphantom{+}0$\\\hline
$S+b_3+x+e_1+e_2+e_4$&$H_{34}^2$&$(1,1,2,2,1)$&$-{1/2}$&$-{1/2}$&$\hphantom{+}0$\\\hline
$S+b_2+x+e_1+e_2+e_5+e_6$&$H_{34}^3$&$(1,1,2,2,1)$&$\hphantom{+}{1/2}$&$\hphantom{+}0$&$-{1/2}$\\\hline
$S+b_1+x+e_3+e_4+e_5+e_6$&$H_{34}^4$&$(1,1,2,2,1)$&$\hphantom{+}0$&$\hphantom{+}{1/2}$&$\hphantom{+}{1/2}$\\\hline
$S+b_1+x+e_4+e_5$&$H_{34}^5$&$(1,1,2,2,1)$&$\hphantom{+}0$&$-{1/2}$&$-{1/2}$\\\hline
$S+b_3+x+z_1$&$H_{13}^1$&$(2,1,2,1,1)$&$-{1/2}$&$-{1/2}$&$\hphantom{+}0$\\\hline
$S+b_3+x+z_1+e_1+e_3+e_4$&$H_{13}^2$&$(2,1,2,1,1)$&$-{1/2}$&$\hphantom{+}{1/2}$&$\hphantom{+}0$\\\hline
$S+b_2+x+z_1+e_2$&$H_{13}^3$&$(2,1,2,1,1)$&$\hphantom{+}{1/2}$&$\hphantom{+}0$&$\hphantom{+}{1/2}$\\\hline
$S+b_2+x+z_1+e_2+e_6$&$H_{14}^1$&$(2,1,1,2,1)$&$\hphantom{+}{1/2}$&$\hphantom{+}0$&$\hphantom{+}{1/2}$\\\hline
$S+b_1+x+z_1+e_3$&$H_{14}^2$&$(2,1,1,2,1)$&$\hphantom{+}0$&$\hphantom{+}{1/2}$&$\hphantom{+}{1/2}$\\\hline
$S+b_1+x+z_1+e_6$&$H_{14}^3$&$(2,1,1,2,1)$&$\hphantom{+}0$&$-{1/2}$&$-{1/2}$\\\hline
$S+b_3+x+z_1+e_3+e_4$&$H_{24}^1$&$(1,2,1,2,1)$&$-{1/2}$&$\hphantom{+}{1/2}$&$\hphantom{+}0$\\\hline
$S+b_3+x+z_1+e_1$&$H_{24}^2$&$(1,2,1,2,1)$&$-{1/2}$&$-{1/2}$&$\hphantom{+}0$\\\hline
$S+b_2+x+z_1+e_1+e_2$&$H_{24}^3$&$(1,2,1,2,1)$&$-{1/2}$&$\hphantom{+}0$&$-{1/2}$\\\hline
$S+b_1+x+z_1+e_3+e_4$&$H_{24}^4$&$(1,2,1,2,1)$&$\hphantom{+}0$&$-{1/2}$&$\hphantom{+}{1/2}$\\\hline
$S+b_1+x+z_1+e_4+e_6$&$H_{24}^5$&$(1,2,1,2,1)$&$\hphantom{+}0$&$\hphantom{+}{1/2}$&$-{1/2}$\\\hline
$S+b_2+x+z_1+e_1+e_2+e_6$&$H_{23}^1$&$(1,2,2,1,1)$&$\hphantom{+}{1/2}$&$\hphantom{+}0$&$\hphantom{+}{1/2}$\\\hline
$S+b_2+x+z_2+e_2+e_5+e_6$&$Z_1$&$(1,1,1,1,8_c)$&$-{1/2}$&$\hphantom{+}0$&$\hphantom{+}{1/2}$\\\hline
$S+b_1+x+z_2+e_3+e_4$&$Z_2$&$(1,1,1,1,8_s)$&$\hphantom{+}0$&$-{1/2}$&$-{1/2}$\\\hline
$S+b_1+x+z_2+e_3+e_5$&$Z_3$&$(1,1,1,1,8_c)$&$\hphantom{+}0$&$-{1/2}$&$\hphantom{+}{1/2}$\\\hline
$S+b_1+x+z_2+e_4+e_6$&$Z_4$&$(1,1,1,1,8_s)$&$\hphantom{+}0$&$-{1/2}$&$-{1/2}$\\\hline
$S+b_1+x+e_5+e_6$&$Z_5$&$(1,1,1,1,8_c)$&$\hphantom{+}0$&$\hphantom{+}{1/2}$&$-{1/2}$\\
\hline
\end{tabular}
}
\caption{\label{tablec}\it Twisted matter spectrum (hidden sector)  and
${SU(2)}^4\times{SO(8)}\times{U(1)}^3$ quantum numbers.}
\end{table}

\section{The superpotential and the top quark Yukawa}

Using the methodology of ref.
\cite{kln} for the calculation of renormalisable and
nonrenormalisable terms, we calculate the cubic level
superpotential of our exophobic Pati--Salam string model.
In particular, we seek to extract models that produce a cubic
level mass term for the heavy generation, but not for the lighter generations,
which should arise from higher order nonrenormalisable terms.
These requirements impose additional non--trivial constraints on the
viable string vacua. Many models
do not produce any coupling of the form ${\bar F}_R F_L h$. Such models do not
admit viable phenomenology as the models should produce at least a top
quark mass term at leading order.
Similarly, models that produce leading mass terms for two or more families
are not viable. The model presented in
ref. \cite{acfkr} produces the cubic level terms
$({\bar F}_{1R}F_{3L}+{\bar F}_{4R}F_{2L})h_3$. In this model therefore
two heavy families may be degenerate in mass.
More appealing are therefore models that produce only a
single mass term at leading order.
The model produced by eq. (\ref{BigMatrix}) is an example of such a model.
The trilevel superpotential is given by
\begin{eqnarray}
&~&
\frac{W_{\rm SM}}{g\,\sqrt{2}}=\nonumber\\
&~&
\bar{F}_{2R}F_{3L}h_1+
\left\{h_1h_1\Phi_{13} + h_2h_2\Phi_{23} + h_3h_3\bar{\Phi}_{23} + h_1h_3\zeta_1\right\} +
\left\{D_1D_2\bar{\Phi}_{12} +
 \right. \nonumber\\
& &
\bar{D_1}D_2\Phi_{12}^{-} + D_1\bar{D_2}\bar{\Phi}_{12}^{-} + \bar{D_1}\bar{D_2}\Phi_{12}
+D_1D_3\bar{\Phi}_{13} +\bar{D_1}D_3\Phi_{13}^{-} +D_1\bar{D_3}\bar{\Phi}_{13}^{-}  +
\nonumber\\
& &
\left.
                                                \bar{D_1}\bar{D_3}\Phi_{13}
+D_2D_3\bar{\Phi}_{23} +\bar{D_2}D_3\Phi_{23}^{-} +D_2\bar{D_3}\bar{\Phi}_{23}^{-}  +
                                               \bar{D_2}\bar{D_3}\Phi_{23}
\right\} +
\nonumber\\
&~&
 \left\{
D_{1}F_{1R}F_{1R} + \bar{D}_{1}\bar{F}_{2R}\bar{F}_{2R}+
D_{2}(\bar{F}_{1R}\bar{F}_{1R} + F_{1L}F_{1L})+
D_{3}(\bar{F}_{4R}\bar{F}_{4R} + F_{2L}F_{2L}) +
\right. \nonumber\\
&~& \left.
                       \bar{D}_{3}(\bar{F}_{3R}\bar{F}_{3R} + F_{3L}F_{3L})+
D_4 ( \bar{F}_{2R}\bar{F}_{3R} + D_2\chi_- +  \bar{D}_2\chi_+ +
D_4\Phi_{13})\right\}+\nonumber\\
&~&
\bar{\Phi }_{13} \chi_-\chi_++\Phi _{23} \bar{\Phi}_{12} \Phi _{13}^-
+\Phi _{13} \bar{\Phi }_{12} \Phi_{23}^-+\Phi _{23} \bar{\Phi }_{13} \Phi _{12}^-
+\Phi_{12}^- \Phi _{23}^- \bar{\Phi }_{13}^- +\nonumber\\
&~&
\Phi _{13}
   \bar{\Phi }_{23} \bar{\Phi }_{12}^-+\Phi _{12}
   \bar{\Phi }_{23} \bar{\Phi }_{13}^-+\Phi _{13}^-
   \bar{\Phi }_{12}^- \bar{\Phi }_{23}^-+\Phi _{12}
   \bar{\Phi }_{13} \bar{\Phi }_{23}^- +
 \nonumber\\
&~&
 \zeta_1{}^2 \bar{\Phi}_{12}^-+\bar{\zeta}_1{}^2 \Phi_{12}^-+
\left(\zeta_3{}^2+\zeta_4{}^2+\zeta_7{}^2\right)
   \bar{\Phi}_{13}^-+\left(\bar{\zeta}_3{}^2+\bar{\zeta}_4{}^2+
\bar{\zeta}_7{}^2\right) \Phi _{13}^- +
    \nonumber\\
&~&
\frac{1}{2} \bar{\zeta}_2
   \bar{\zeta}_5 \chi_++\zeta_2{}^2 \bar{\Phi }_{12}+\left(\zeta_5{}^2+
\zeta_6{}^2\right) \bar{\Phi }_{23}+\Phi _{12}
   \bar{\zeta}_2{}^2+\Phi _5 \left(\zeta_1 \bar{\zeta}_1+\zeta_2
   \bar{\zeta}_2\right) +
    \nonumber\\
&~&
\Phi _2 \left(\zeta_5 \bar{\zeta}_5+\zeta_6
   \bar{\zeta}_6\right)+\Phi _{23}
   \left(\bar{\zeta}_5{}^2+\bar{\zeta}_6{}^2\right)+\Phi _4 \zeta_7
   \bar{\zeta}_7+\frac{\zeta_4 \zeta_5
   \bar{\zeta}_2}{\sqrt{2}}+\frac{\zeta_2 \bar{\zeta}_3
   \bar{\zeta}_5}{\sqrt{2}}
\end{eqnarray}

The string vacuum contains three anomalous $U(1)$s
\begin{equation}
{\rm Tr}{U(1)}_1=-12~~;~~
{\rm Tr}{U(1)}_2=-24~~;~~
{\rm Tr}{U(1)}_3=-12
\end{equation}
redefining we obtain two anomaly-free
\begin{eqnarray}
{U(1)}'_1={U(1)}_1-{U(1)}_3\label{u1p}\\
{U(1)}'_2={U(1)}_1-{U(1)}_2+{U(1)}_3\label{u2p}
\end{eqnarray}
and one anomalous combination
\begin{eqnarray}
{U(1)}'_A={U(1)}_1+2\,{U(1)}_2+{U(1)}_3\ ,\ {\rm Tr}{U(1)}_A=-72\label{upa}
\end{eqnarray}

The electroweak Higgs doublets come in pairs and are accommodated
in the Pati--Salam bi-doublets $h_1,h_2,h_3$. Their mass matrix is
\begin{eqnarray}
M_h\sim\bordermatrix{
&h_1&h_2&h_3\cr
h_1&\Phi_{13}&\frac{\zeta_1}{\sqrt{2}}&0\cr
h_2&\frac{\zeta_1}{\sqrt{2}}&\bar{\Phi}_{23}&0\cr
h_3&0&0&\Phi_{23}}
\end{eqnarray}
In order to keep $h_1$ massless we need to impose the condition
\beq
\Phi_{13}\,\bar{\Phi}_{23}-\frac{\zeta_1^2}{2}=0.
\label{doubletconstraint}
\eeq

Next, we discuss the colour--triplet mass matrix in our string derived
Pati--Salam model. Three pairs of colour--triplets arise in the model from
the untwisted Neveu--Schwarz sector, and are accommodated in the sextet of the
Pati--Salam $SU(4)$. we denote these by
$D_i=d_i(3,1,1)+d^c_i(\bar{3},1,1)$,
$\bar{D}_i=\bar{d}_i({\bar 3},1,1)+\bar{d}^c_i({3},1,1)$.
An additional sextet arises in the model from a twisted sector.
A further pair of colour triplets is obtained from
the heavy Higgs states,
$\bar{F}_{1R}$ and $F_{1R}$ that are used to break the Pati--Salam
symmetry, and must get a VEV of the order of the GUT scale.
We denote the colour triplets in these fields by
$F_{\alpha R}=d_{\alpha H}+\dots$. At the cubic level the colour triplet mass
matrix then takes the form,
\begin{eqnarray}
M_D=\bordermatrix{
&d_1&d_2&d_3&\bar{d}_1&\bar{d}_2&\bar{d}_3&d_4&d_{1H}\cr
d^c_1&0 & \bar{\Phi }_{12} & \bar{\Phi }_{13} & 0
& \bar{\Phi }_{12}^- & \bar{\Phi }_{13}^- & 0 & F_{1R}\cr
d^c_2&\bar{\Phi }_{12} & 0 & \bar{\Phi }_{23} & \Phi _{12}^-
& 0 & \bar{\Phi }_{23}^- & \chi_- & 0 \cr
d^c_3&\bar{\Phi }_{13} & \bar{\Phi }_{23} & 0 & \Phi _{13}
& \Phi _{23}^- & 0 & 0 & 0 \cr
\bar{d}^c_1&0 & \Phi _{12}^- & \Phi _{13} & 0 & \Phi _{12}
& \Phi _{13}  &0& 0 \cr
\bar{d}^c_2&\Phi _{12}^- & 0 & \Phi _{23}^- & \Phi _{12} & 0
& \Phi _{23} & \chi_+ & 0 \cr
\bar{d}^c_3&\bar{\Phi }_{13}^- & \bar{\Phi }_{23}^- & 0
& \Phi _{13} & \Phi _{23} & 0 & 0 & 0 \cr
d^c_4&0 & \chi_- & 0 & 0 & \chi_+ & 0 & \Phi _{13} & 0 \cr
\bar{d}^c_{1H}&0 & \bar{F}_{1R} & 0 & 0 & 0 & 0 & 0 & 0}
\label{tmm}
\end{eqnarray}
We have ${\rm det}(M_D)\sim \Phi_{13}^2$ so in order to keep
triplets heavy and $h_1$ light we need
$\{\Phi_{13}, \zeta_1,\bar{\Phi}_{23}\}\ne0$.

Next, we examine the pattern of symmetry breaking. The anomalous $U(1)_A$
is broken by the Green--Schwarz--Dine--Seiberg--Witten mechanism \cite{dsw}
in which a potentially large Fayet--Iliopoulos $D$--term
$\xi$ is generated by the VEV of the dilaton field.
Such a $D$--term would, in general, break supersymmetry, unless
there is a direction $\hat\phi=\sum\alpha_i\phi_i$ in the scalar
potential for which $\sum Q_A^i\vert\alpha_i\vert^2<0$ and that
is $D$--flat with respect to all the non--anomalous gauge symmetries
along with $F$--flat. If such a direction
exists, it will acquire a VEV, cancelling the Fayet--Iliopoulos
$\xi$--term, restoring supersymmetry and stabilising the vacuum.
Assuming VEVs for the non-Abelian gauge singlets and a pair of PS breaking Higgs,
$F_{1R}=\bar{F}_{1R}=M_G$,
the $D$--flatness constraints in our model are given by:
\begin{eqnarray}
{U(1)}'_1 &:& \left(\left|\Phi_{12}\right|^2-\left|\bar{\Phi}_{12}
\right|^2\right)+
\left(\left|\Phi_{12}^{-}\right|^2-\left|\bar{\Phi}_{12}^{-}\right|^2\right)
+2\left(\left|\Phi_{13}^{-}\right|^2-\left|\bar{\Phi}_{13}^{-}\right|^2\right)
\nonumber\\
&~&-\left(\left|\Phi_{23}\right|^2-\left|\bar{\Phi}_{23}\right|^2\right)
+\left(\left|\Phi_{23}^-\right|^2-\left|\bar{\Phi}_{23}^-\right|^2\right)
+\frac{1}{2}\sum_{i=1,2}\left(\left|\zeta_{i}\right|^2-\left|\bar{\zeta}_{i}\right|^2\right)
\nonumber\\
&~&
-\frac{1}{2}\sum_{i=5,6}\left(\left|\zeta_{i}\right|^2-\left|\bar{\zeta}_{i}
\right|^2\right)
+\sum_{i=3,4,7}\left(\left|\zeta_{i}\right|^2-\left|\bar{\zeta}_{i}\right|^2\right)-\frac{1}{2}\left|{F}_{1R}\right|^2
=0\label{du1}
\end{eqnarray}
\begin{eqnarray}
{U(1)}'_2 &:&
2\left(\left|\Phi_{12}^{-}\right|^2-\left|\bar{\Phi}_{12}^{-}\right|^2\right)
+2\left(\left|\Phi_{13}\right|^2-\left|\bar{\Phi}_{13}\right|^2\right)
-2\left(\left|\Phi_{23}^-\right|^2-\left|\bar{\Phi}_{23}^-\right|^2\right)\nonumber\\
&~&+
\left(\left|\zeta_{1}\right|^2-\left|\bar{\zeta}_{1}\right|^2\right)+
2\left|\chi_{-}\right|^2+\frac{1}{2}\left(\left|\bar{F}_{1R}\right|^2-\left|F_{1R}\right|^2\right)=0\label{du2}
\\
{U(1)}'_A &:&
3\left(\left|\Phi_{12}\right|^2-\left|\bar{\Phi}_{12}\right|^2\right)-
\left(\left|\Phi_{12}^{-}\right|^2-\left|\bar{\Phi}_{12}^{-}\right|^2\right)
+2\left(\left|\Phi_{13}\right|^2-\left|\bar{\Phi}_{13}\right|^2\right)
\nonumber\\
&~&+
3\left(\left|\Phi_{23}\right|^2-\left|\bar{\Phi}_{23}\right|^2\right)
+\left(\left|\Phi_{23}^-\right|^2-\left|\bar{\Phi}_{23}^-\right|^2\right)
-\frac{1}{2}\left(\left|\zeta_{1}\right|^2-\left|\bar{\zeta}_{1}\right|^2\right)
\nonumber\\
&~&
+\frac{3}{2}\sum_{i=2,5,6}\left(\left|\zeta_{i}\right|^2-
\left|\bar{\zeta}_{i}\right|^2\right)
+3\left|\chi_{+}\right|^2-\left|\chi_{-}\right|^2
\nonumber\\
&~&
-\frac{1}{2}\left|F_{1R}\right|^2-\left|\bar{F}_{1R}\right|^2=
+\frac{3\,g^2}{16\pi^2}M^2\equiv\xi. \label{duA}
\end{eqnarray}
In eq. (\ref{duA}) $g$ is the gauge coupling in the effective field theory,
and $M$ is the so--called reduced Planck mass
$M\equiv M_{\rm Planck}/\sqrt{8\pi}$. In setting $\xi$ we followed
the conventions of \cite{cew}.
The set of $F$--flatness constraints are obtained by requiring
\beq
\langle F_i\equiv
{{\partial W}\over{\partial\eta_i}}\rangle=0\label{fterms}
\eeq
where $\eta_i$ are all the fields that appear in the model.
The solution ({\it i.e.}\  the choice of fields
with non--vanishing VEVs) to the set of
equations (\ref{du1})--(\ref{fterms}),
though nontrivial, is not unique. Therefore in a typical model there exist
a moduli space of solutions to the $F$ and $D$ flatness constraints,
which are supersymmetric and degenerate in energy \cite{moduli}.
Assuming VEVs for the non-Abelian gauge singlets and a pair of PS breaking Higgs,
$F_{1R}=\bar{F}_{1R}=M_G$, the following 9 parameter exact solution
\beq
\left\{\Phi _3,\Phi _4,\Phi _6,\bar{\Phi }_{23},\Phi_{23}^-,
\bar{\Phi }_{23}^-,\Phi _{13}^-,\bar{\Phi}_{13}^-,
\bar{\Phi }_{12}\right\}
\label{freeparameters}
\eeq
satisfies all $F$-flatness equations while keeping one linear
combination of the bi-doublets ($h_1,h_2$) massless:
\begin{eqnarray}
0&=&\Phi_1=\Phi_2=\chi_+=\chi_-=\zeta_i=
\bar{\zeta}_i,i=3,\dots,7\label{fphi12zeta37}\\
\Phi_5&=&-\frac{2i}{\sqrt{3}}\frac{\bar{\Phi}_{12}}{\bar{\Phi}_{23}}
\sqrt{\frac{\Phi_{13}^- \Phi_{23}^- \bar{\Phi}_{23}^-}{\bar{\Phi}_{13}^-}}
\label{fphi5}\\
\Phi_{23}&=&\frac{\Phi_{23}^-\bar{\Phi}_{23}^-}{\bar{\Phi}_{23}} 
~~~~~~~~~~~~~~~~~~~~~~,~~~~
\Phi_{13} = -\frac{\Phi_{13}^- \bar{\Phi}_{23}^-}{3\bar{\Phi}_{23}}\label{fphi13}\\
\bar{\Phi}_{13}&=&-\frac{3 \bar{\Phi }_{23}
\bar{\Phi }_{13}^-}{\bar{\Phi }_{23}^-} ~~~~~~~~~~~~~~~~~~,~~~~
\Phi_{12} = -\frac{\bar{\Phi }_{12} \Phi _{13}^- \Phi _{23}^- \bar{\Phi
   }_{23}^-}{3 \bar{\Phi }_{23}{}^2 \bar{\Phi }_{13}^-}\label{fphi12}
\end{eqnarray}
\begin{eqnarray}
\Phi_{12}^-&=&\frac{\bar{\Phi }_{12} \Phi _{13}^- \bar{\Phi }_{23}^-}{3
   \bar{\Phi }_{23} \bar{\Phi }_{13}^-} ~~~~~~~~~~~~~~~~~~,~~~~
\bar{\Phi}_{12}^- = -\frac{\bar{\Phi }_{12} \Phi _{23}^-}{\bar{\Phi }_{23}}\\
\zeta_1&=&i \sqrt{\frac{2\Phi _{13}^- \bar{\Phi }_{23}^-}{3}}  ~~~~~~~~~~~~~~,~~~~
~\bar{\zeta}_1 = - \sqrt{2 \Phi _{23}^- \bar{\Phi }_{13}^-}\\
\zeta_2&=&i \sqrt{\frac{2\Phi _{23}^-
   \Phi _{13}^- \bar{\Phi }_{23}^-}{3\bar{\Phi }_{23}}} ~~~~~~~~~~~~~,~~~~
~\bar{\zeta}_2 = \sqrt{2 \bar{\Phi }_{23} \bar{\Phi }_{13}^-}\label{fzeta2}
\end{eqnarray}

The triplet mass matrix \eqref{tmm} determinant is
\begin{eqnarray}
\det{M_D}=  -\frac{64}{27}\frac{F_{1R}
\bar{F}_{1R}\bar{\Phi }_{12} \Phi _{13}^-{}^3
   \Phi _{23}^-{}^2 \bar{\Phi }_{23}^-{}^3}{\bar{\Phi}_{23}{}^3}
\end{eqnarray}
and thus all triplets are massive.

\begin{figure}[!ht]
\centering
\includegraphics[width=10cm]{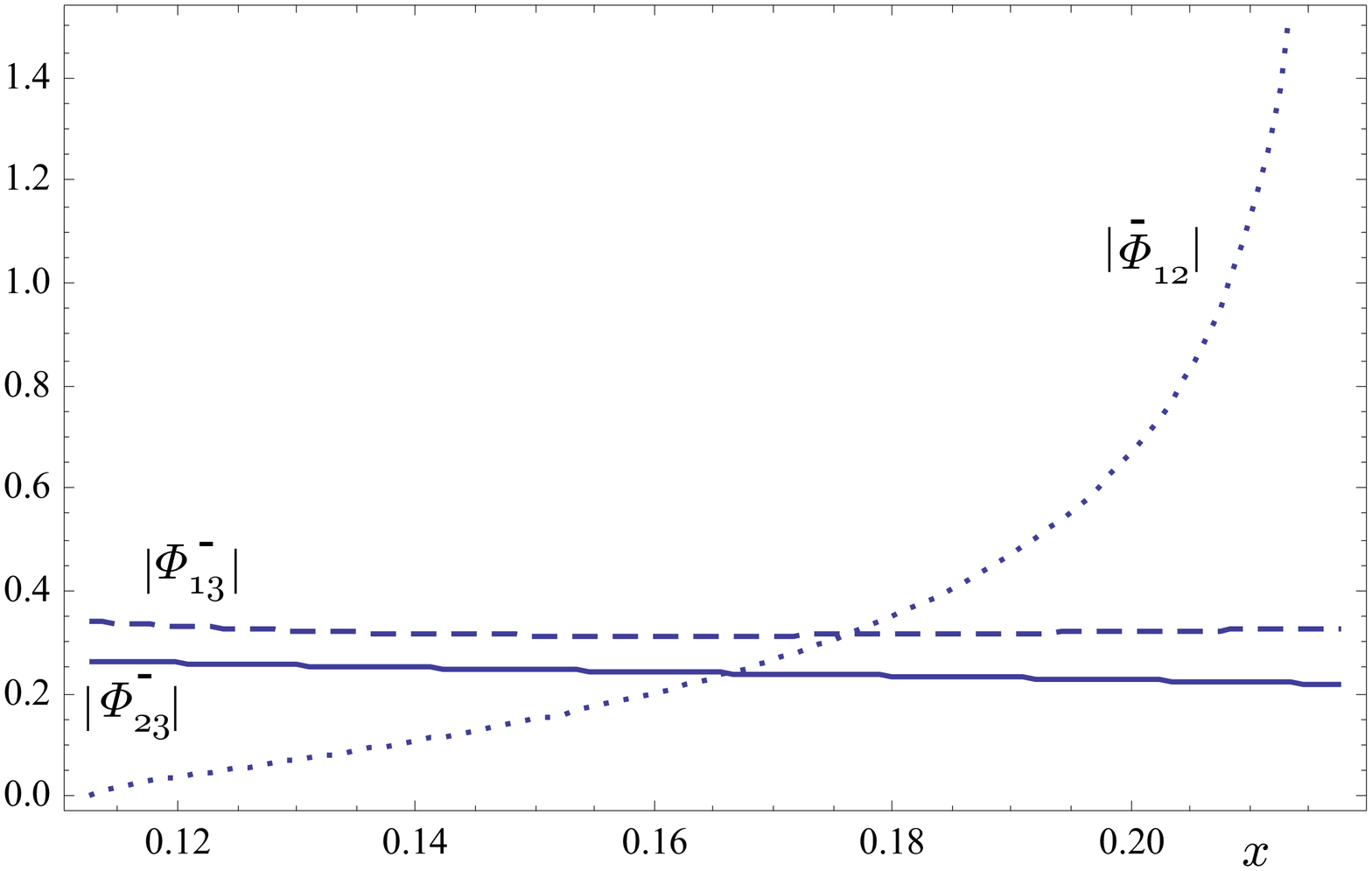}
\caption{\label{sold}
\it Solution of of the $D$--flatness equations for
$\vert {\Phi}_{13}^- \vert$,  $\vert {\bar\Phi}_{23}^- \vert$
and
$\vert {\bar\Phi}_{12}^- \vert$  as a
function of $\chi= \vert {\bar\Phi}_{23}\vert=\vert {\bar\Phi}_{13}^-\vert =
{1\over2} \vert {\bar\Phi}_{23}^- \vert$ (all VEVs are in units of  $\sqrt{\xi}$).}
\end{figure}

\begin{figure}[!ht]
\centering
\includegraphics[width=12cm]{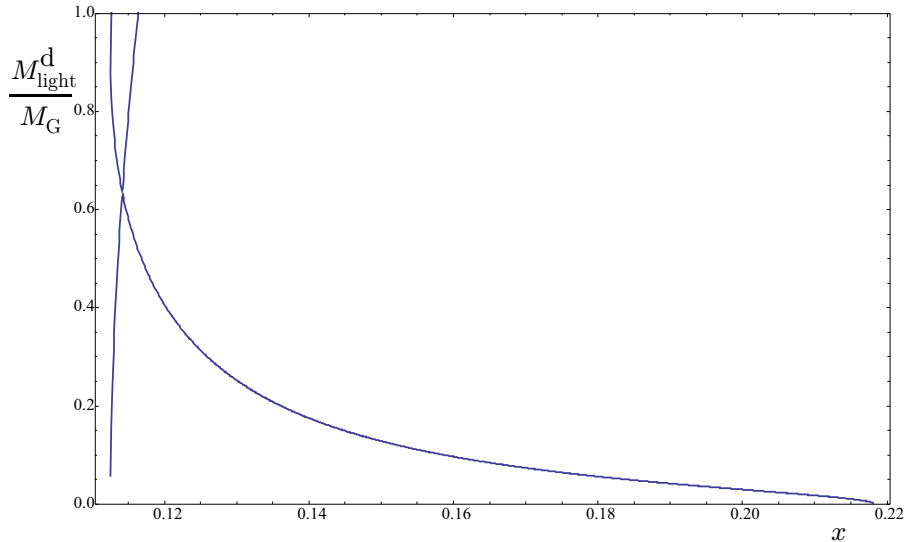}
\caption{\label{mdlight}
\it The ratio  of the two lightest colour triplet mass over $M_{\rm GUT}$ as a
function of $\chi= \vert {\bar\Phi}_{23}\vert=\vert {\bar\Phi}_{13}^-\vert =
{1\over2} \vert {\bar\Phi}_{23}^- \vert$ (in units of  $\sqrt{\xi}$). }
\end{figure}

For this $F$--flatness solution, the  three $D$--flatness equations (\ref{du1}--\ref{duA}) depend on seven
parameters,
$\vert {\bar\Phi}_{23}\vert,~
\vert {\Phi}_{23}^-\vert,~
\vert {\bar\Phi}_{23}^-\vert,~
\vert {\Phi}_{13}^-\vert,~
\vert {\bar\Phi}_{13}^-\vert,~
\vert {\bar\Phi}_{12}\vert,$
and $\vert F_{1R}\vert = \vert {\bar F}_{1R}\vert.$
Setting $\vert F_{1R}\vert = \vert {\bar F}_{1R}\vert = M_{\rm G}=0.02 \sqrt{\xi}$
the $D$--flatness equations can be solved numerically in terms of three
parameters. Choosing, for example,
$\vert {\bar\Phi}_{23}\vert=\vert {\bar\Phi}_{13}^-\vert =
{1\over2} \vert {\bar\Phi}_{23}^- \vert = \chi$
we can solve numerically for
$\vert {\Phi}_{13}^- \vert$,  $\vert {\bar\Phi}_{23}^- \vert$
and
$\vert {\bar\Phi}_{12}^- \vert$.
The results are shown in figure \ref{sold}.
In figure \ref{mdlight} we plot the mass of the two lightest  colour triplets for
the one parameter solution displayed in figure \ref{sold}.
From the figure we note that for singlet VEVs of the order of $0.1\sqrt{\xi}$
the lightest triplet mass is of the order of $0.4M_{\rm GUT}$. Thus the additional colour
triplets are heavy enough to protect proton from decaying through dangerous  triplet mediated  dim-5
operators \cite{lpd}. Additionally, we note that the three $U(1)$ symmetries
in eqs. (\ref{u1p}, \ref{u2p}, \ref{upa}) are broken in the $F$-- and $D$--flat vacuum.

\section{Conclusions}\label{conclude}

In this paper we analysed the phenomenology of an exemplary exophobic
Pati--Salam heterotic string vacuum, in which exotic fractionally
charged states exist in the massive spectrum, but not among the
massless states. In that respect the exophobic models are distinguished
from other models in which exotic states gain heavy mass
by vacuum expectation values of Standard Model singlet fields.
Our exophobic model also contains the
Higgs representations that are needed to break the gauge symmetry
to that of the Standard Model and to generate fermion masses
at the electroweak scale. One can then start to probe the
phenomenology of such models in more detail. We showed
in particular that the presence of a top
quark Yukawa coupling at leading order places
an additional strong constraint on the viability of the models.
In many models a top quark Yukawa may not exist at all, whereas
in others two or more generations may obtain a mass term at leading
order. In our exemplary model a mass term at leading
order exist only for one family. Additionally,
we demonstrated the existence of supersymmetric $F$-- and $D$--flat
directions that give heavy mass to all the
colour triplets beyond those of the Standard Model
and leave one pair of electroweak Higgs doublets light.
Hence, below the Pati--Salam breaking scale the spectrum of our
model consists solely of that of the Minimal Supersymmetric
Standard Model. We remark that while there exist other models
in which the exotic states are decoupled along flat directions,
in many of these models the mass scale of the exotic states
is ambiguous as the relevant mass terms arise from higher
order superpotential terms that are expected to be suppressed
compared to the leading string scale mass terms \cite{raby}.
The novelty in our model is that the exotic states are
absent from the massless spectrum to begin with and
hence necessarily have string scale masses. In this respect the model
is superior to earlier constructions. Further analysis of higher
order terms in the superpotential can now be pursued to
confront the model with the detailed Standard Model mass
and mixing data. We note that the interplay between statistical
searches and detailed analysis of specific models takes us a step further
toward the construction of string models that reproduce the phenomenological
Standard Model data. We will return to these issues in future publications.

\section{Acknowledgements}

AEF would like to thank the University of Oxford for hospitality.
AEF is supported in part by STFC under contract PP/D000416/1.
JR work is supported in part by the EU under contract
PITN-GA-2009-237920.



\bigskip
\medskip

\bibliographystyle{unsrt}

\end{document}